\documentclass[journal]{IEEEtran}

\pdfoutput=1

\usepackage[utf8]{inputenc}
\usepackage[T1]{fontenc}

\usepackage[english]{babel}

\usepackage{lipsum}

\usepackage{color}
\usepackage{amsmath}

\usepackage{graphicx, subfigure} 
\usepackage[
    backend=biber,
    sorting=none,
    style=numeric,
    maxnames=99
  ]{biblatex}
 
\begin{document}

\title{On Modeling Network Slicing Communication Resources with SARSA Optimization}


\author{ Eduardo S. Xavier, Nazim Agoulmine and Joberto S. B. Martins
\IEEEcompsocitemizethanks{


\IEEEcompsocthanksitem Xavier, Eduardo F. is with Salvador University (UNIFACS), Brazil - eduardo.sidney@animaeducacao.com.br
\IEEEcompsocthanksitem Agoulmine, Nazim is with University of Paris Saclay, University of Evry - IBISC Lab, France - nazim.agoulmine@univ-evry.fr
\IEEEcompsocthanksitem Martins, Joberto S. B. is with Salvador University (UNIFACS), Brazil - joberto.martins@gmail.com
}

\thanks{This work IS supported by ANIMA Institute and FAPESP - MCTIC - Project 2018/23097-3.}}

\maketitle

\markboth{ArXiv Preprint - FEBRUARY 2023}{
Bhardwaj \MakeLowercase{\textit{et al.}}: Skeleton of IEEEtran.cls for Journals in VIM-Latex}



\begin{abstract}
Network slicing is a crucial enabler to support the composition and deployment of virtual network infrastructures required by the dynamic behavior of networks like 5G/6G mobile networks, IoT-aware networks, e-health systems, and industry verticals like the internet of vehicles (IoV) and industry 4.0. The communication slices and their allocated communication resources are essential in slicing architectures for resource orchestration and allocation, virtual network function (VNF) deployment, and slice operation functionalities. The communication slices provide the communications capabilities required to support slice operation, SLA guarantees, and QoS/ QoE application requirements. Therefore, this contribution proposes a networking slicing conceptual model to formulate the optimization problem related to the sharing of communication resources among communication slices. First, we present a conceptual model of network slicing, we then formulate analytically some aspects of the model and the optimization problem to address. Next, we proposed to use a SARSA agent to solve the problem and implement a proof of concept prototype. Finally, we present the obtained results and discuss them.


\end{abstract}




\section{Introduction} \label{sec:Introduction}

Network slicing is a crucial enabler to support the composition and deployment of virtual network infrastructures required by the dynamic behavior of networks like 5G/6G mobile networks, IoT-aware networks, e-health systems, and industry verticals like the internet of vehicles (IoV) and industry 4.0 \cite{wijethilaka_survey_2021} \cite{barakabitze_5g_2020} \cite{yousaf_network_2017}. In general, the slicing process results from the need to share resources among existing infrastructures to improve performance, provide cost-efficient solutions, and optimize operation \cite{thiruvasagam_resilient_2021}.

This technology is already used in the context of 5G networks \cite{wijethilaka_survey_2021} \cite{chahbar_comprehensive_2021} and provided  as a service (slice-as-a-Service: SlaaS) by network operators. This allows customs to create their private  virtual networks (slices) tailored to their specific application domains and to develop their own business models. Network slicing is expanding its use in other scenarios of telecommunication networks, content provider networks (ISPs), experimental networks, and IoT systems, among others \cite{kovacevic_multi-domain_2020}.
 
Network slice instance life cycle process such as commissioning, operating, and decommissioning \cite{wijethilaka_survey_2021} requires appropriate network communication resources. A communication slice \footnote{A specialized slice that provides communication services among network slicing entities} eventually represents a set of communication resources that can be used in the slicing process. It holds resources like links, optical slots, virtual private networks (VPNs), and other communication facilities necessary to provide the exchange of information among logical slices, and architectural slicing entities and for supporting the slicing process functionalities.

The communication slice resources significantly impact the performance of the resulting sliced virtual network (SVN) or virtual network operator (VNO). Among the most common network characteristics that impact the network slicing process, we can mention delay-aware network slicing like in 5G deployments \cite{prados-garzon_asynchronous_2021}, quality of service (QoS) aware network slicing \cite{yousaf_network_2017}, energy-aware network slicing \cite{xiao_dynamic_2018}, and, in general, application-dependent and multi-domain network slicing \cite{samdanis_network_2016}.


The objective of this paper is therefore to propose a conceptual model of slice communication and formulate analytically some of its aspects. The model should be able to capture the set of communication resources to support the optimization of the allocation of communication resources to the different slices on top of various underlying technologies (e.g. Elastic Optical Networks - EON \cite{duraes_evaluating_2017}, MultiProtocol Label Switching - MPLS \cite{de_ghein_mpls_2007}, others)



 


This paper is organized as follows. Section \ref{sec:RelatedWork} presents the related work and Section \ref{sec:SVN} introduces the concept of multidomain sliced virtual networks. Section \ref{sec:Model} presents a conceptual and analytical model for a communication slice used in the network slicing process. Section \ref{sec:ProofOfConcept} presents a proof of concept of using the models with a SARSA agent optimizing the allocation of bandwidth resources  for a communication slice. Finally, Section \ref{sec:Conclusion} presents the final considerations.


\section{Related Work} \label{sec:RelatedWork}

 There have been  a very significant number of state-of-art research projects launched in the area during the last decade such as SFI2 (Slicing Future Internet Infrastructures) \cite{dias_sfi2_2019} \cite{martins_sfi2_2022}, NECOS (Novel Enablers for Cloud Slicing) \cite{clayman_necos_2021}, SELFNET \cite{DBLP:journals/ijdsn/NightingaleWCCU16} and MATILDA \cite{gouvas_design_2017}, standardization initiatives launched by the IETF (Internet Engineering Task Force) \cite{ietf_framework_2021}, 3GPP (3rd Generation Partnership Project) \cite{3gpp_5g-evolution-3gpp_2020}, ITU (ITU-T - Telecommunication Standardization) \cite{itu-t_framework_2012}, ETSI (European Telecommunications Standards Institute) \cite{etsi_mobile_2015} and ONF (Open Networking Foundation) \cite{onf_-_open_networking_foundations_applying_2016} and published surveys \cite{barakabitze_5g_2020} \cite{zhang_overview_2019} \cite{foukas_network_2017} \cite{kaloxylos_survey_2018} \cite{afolabi_network_2018}. These different initiatives have focused on different technical aspects, architectures, and slicing strategies, and all require communication slices to operate and manage the provided functionalities.
 
 

However, these  slicing architectures, projects, and initiatives did only address the conceptual and analytical modeling of the basic structures and functionalities that compose the slicing process in a preliminary way or did only indicate them as future challenges to solve. To the best of our knowledge, the conceptual and analytical modeling of communication slices is a new contribution to the network slicing domain.

 

\section{Resources, Slice and Sliced Virtual Network (SVN)} \label{sec:SVN}

\begin{figure}[htbp]
\begin{center}
\includegraphics[scale=0.45]{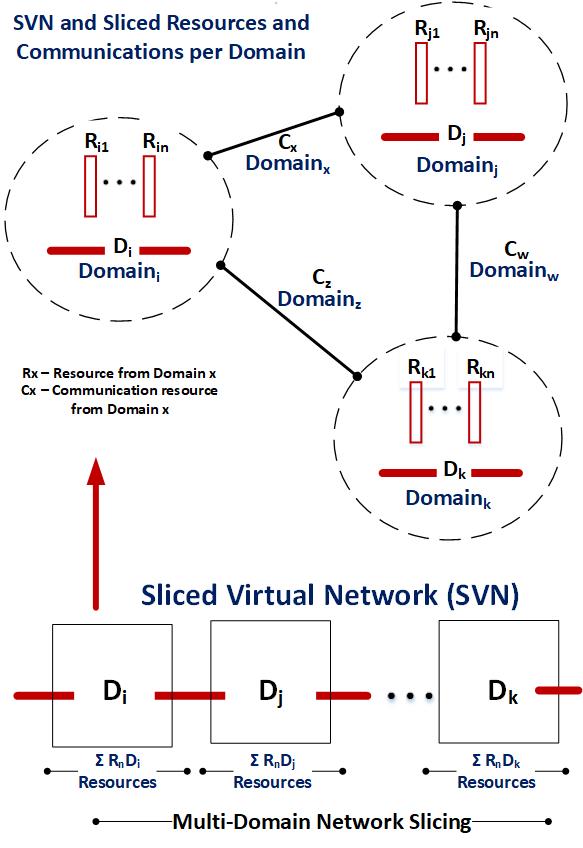}
\caption{A Multi-Domain Sliced Virtual Network (SVN) and its Resources}
\label{fig:SVNResources}
\end{center}
\end{figure}

A multi-domain Sliced Virtual Network (SVN) as viewed in Figure \ref{fig:SVNResources} is a multi-domain or a multi-tenant\footnote{For the scope of this paper, a tenant can be a network domain, a service provider, a business unit, or a specific multi-tier or single-application tier providing resources for network slicing.} infrastructure that is dynamically configured and deployed by requesting and orchestrating resources from a pool of providers on domains.

\subsection{The Slice}

For the scope of this paper aiming at the slicing model and deployment understanding, it is essential to conceptualize the vision of a \textit{slice} as a component of the sliced virtual network.

We define a slice as a specific resource, service, function, or set of resources, services, and functions virtualized, shared, and grouped using any software or hardware facility. The slice with its resources, services, and functions physically resides in nodes or other physical deployments in domains.

As such, slice resource examples are virtual machines, virtual switches with hosts deployed with OpenFlow, chunks of bandwidth belonging to a physical link, slots of a fiber EON deployment, LSP MPLS connections, shared spectrum in 5G radio access networks (RAN), and others. Slice function and service examples are virtual network functions (VNFs) deployed over a network providing specific services or facilities to the user.

\begin{figure}[htbp]
\begin{center}
\includegraphics[scale=0.45]{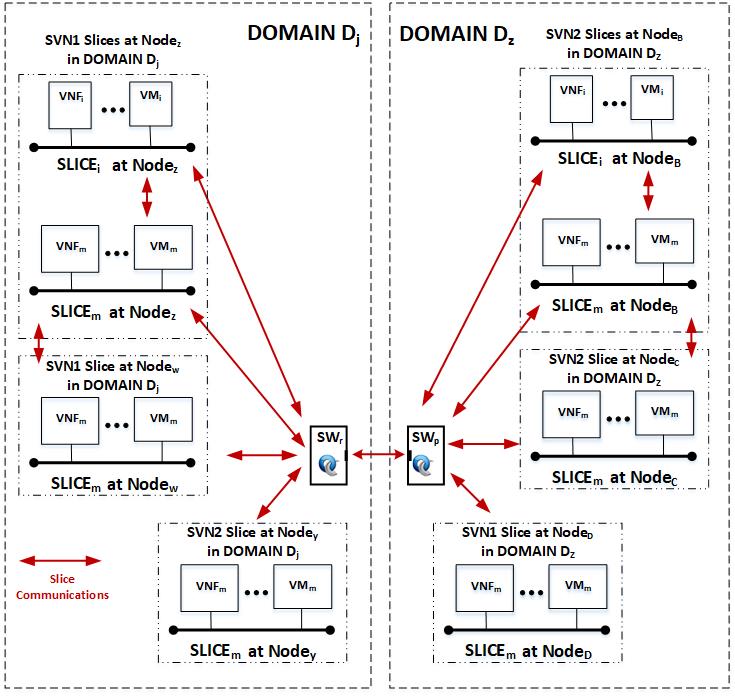}
\caption{A Generic SVN with Slices View}
\label{fig:SLICE}
\end{center}
\end{figure}

Considering this slice basic concept, an SVN encompasses resources, services, and functions with the necessary communication resources to interconnect them inside domains and between domains as illustrated in Figure \ref{fig:SLICE}.

In general, resources belonging to the same SVN reside in different domains and are physically or virtually attached to nodes in their respective domains (Figure \ref{fig:SLICE}).

The network slicing architecture functionalities (resource marketplace, resource broker, resource orchestrator, slice instantiation, slice monitoring, and others) are distributed in terms of the domains participating in the SVN deployment and certainly, depend on the proposed architecture and the deployed functional blocks of the network slicing architecture (SELFNET, NECOS, SFI2, MATILDA, other).


\begin{figure}[htbp]
\begin{center}
\includegraphics[scale=0.4]{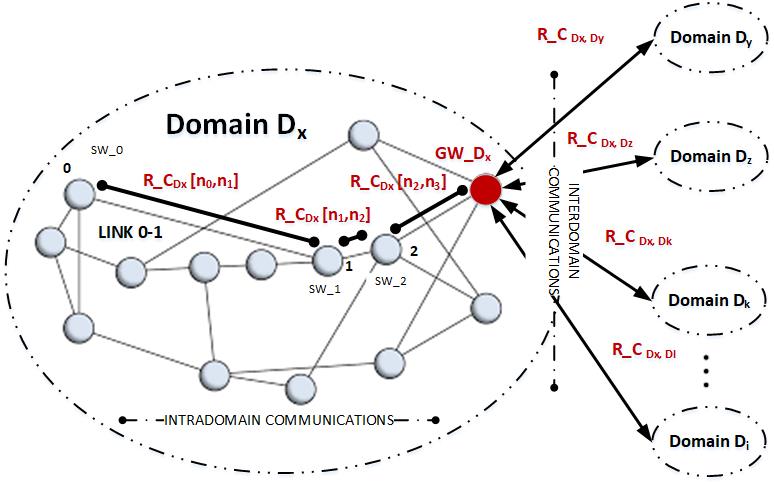}
\caption{Intradomain and Interdomain Communication Slices}
\label{fig:CommunicationSlices}
\end{center}
\end{figure} 

\subsection{Communication Resources and Communication Slice}

In order to allow the execution of the network slicing process and functionalities in any deployed slicing architecture, it is necessary to allocate communication resources allowing communication among the entities involved in the slicing process. Furthermore, once the SVN is deployed, communication resources are also necessary to support the communication requirements of the applications running (slice operation).


The generic view of communication resources used by a network slicing infrastructure to enable resource orchestration, deployment, and slice operation is illustrated in Figure \ref{fig:CommunicationSlices}.

We assume that the slicing process to create a sliced virtual network (SVN) involves single or multiple domains ($D_x, ..., D_z$). Each domain is generically configured by a single or a set of nodes ($n_i, ... n_j$) hosting resources and domains that are interconnected by communication resources.


A \textit{communication slice} is then defined as a set of communication resources orchestrated and allocated between slices, nodes, network-slicing entities, and domains. As such, the domain nodes ($n_i, ... n_j$) hosting resources and domains are interconnected by communication slices ($C_x, ... C_y$).



We identify two types of communication slices that are orchestrated and deployed with distinct configurations and characteristics:

\begin{itemize}
    \item Intradomain communication slices; and
    \item Interdomain communication slices.
\end{itemize}

In infrastructures composed of network domains, the modeling assumes that a gateway concentrates all communications between different domains.

We focus in this paper specifically on interdomain communications and how to model it in terms of communication slices.



\section{Network Slicing Interdomain Communications} \label{sec:Model}


The objective of a network slicing interdomain communication model is to formally structure and capture the needs in terms of communications for the slicing process. It also allows the identification of parameters leading to the optimization of the resource allocation process.

\subsection{Network Slicing Assumptions}

We first introduce the following assumptions in the context of network-slicing interdomain communications that are necessary for our modeling and problem formulation:

\begin{itemize}
    \item Each network domain is SDN-compatible;
    \item Each network domain gateway $GW\_D_i$ (Figure \ref{fig:CommunicationSlices}) is an SDN-enabled switch whose programmed behavior is to route packets between domains;
    \item Each network domain implements monitoring mechanisms to collect performance monitoring parameters;
    \item All intradomain and interdomain links are configurable in terms of allocated resources; and
    \item All network domains support network resource identification and has capabilities for resource allocation.

\end{itemize}



\subsection{Network Slicing Model}

Based on these assumptions, we can now specify an analytical model of multi-domain SVN considering a set of network domains federating together their resources and infrastructures to the slicing process:

\begin{eqnarray}\label{eq:PredictNextQ2}
\aleph = <D_i^{l_i}, D_j^{l_j},D_k^{l_k},...,D_z^{l_z}>
\end{eqnarray}

Where:
\begin{itemize}
    \item $D_i^{l_i}$ is a network infrastructure domain located at site $l_i$.
\end{itemize}

\begin{table}[]
\begin{tabular}{ll}
\hline
Notation & Description \\ \hline

$D_i^{l_i}$                         & The domain $i$ located in physical location $l_i$         \\
$RD_i^{l_i}$                        & Domain's set of shareable resources at a physical location   \\
$R_i^{{D_i^{l_i}}}$                 & A shareable resource at domain $D_i^{l_i}$ \\
$R\_IS_{D_i}^{l_i}$                 & The infrastructure and service resources  \\
$R\_C_{D_i}^{l_i}$                  & The network communication resources  \\
$B_{D_i,D_j}$                       & Bandwidth between domains \\
$L_{D_i,D_j}$                       & Packet loss between domains    \\
$Dl_{D_i,D_j}$                      & Delay between domains   \\
$B_{n_i,n_j}$                       & Bandwidth between nodes    \\
$L_{n_i,n_j}$                       & Packet loss between nodes    \\
$Dl_{n_i,n_j}$                      & Delay between nodes   \\
$P\_RC_{D_k,D_j}^{l_i}$             & Set of communication's link parameters between domains    \\

\hline
\end{tabular}
\caption{Notation and variables}
\label{tab:parameters}
\end{table}

Each network infrastructure domain  $D_i^{l_i}$ has a set of shareable resources such as:

\begin{eqnarray}\label{eq:PredictNextQ2}
RD_i^{l_i} = <R_i^{{D_i^{l_i}}}, R_j^{{D_i^{l_i}}},R_k^{D_i^{l_i}},...,R_z^{D_i^{l_i}}>
\end{eqnarray}

Where:
\begin{itemize}
    \item $RD_i^{l_i}$ is the set of shareable resources provided by $D_i$ and located at site $l_i$;
    \item $R_i^{{D_i^{l_i}}}$ is one particular shareable resource.
\end{itemize}

There are different types of resources at each network infrastructure domain location $D_i^{l_i}$:
\begin{itemize}
    \item Infrastructure appliance like virtual machines, access points, and IoT devices;
    \item Computing services like virtual network functions (VNF), storage and computing services; and
    \item Communications services like physical links, LSPs (MPLS  Link Switched Paths), fiber lambdas, and 5G connections.
\end{itemize}

For the purpose of the SVN model, we distinguish between two types of resources:
\begin{itemize}
    \item Infrastructure and service resources - $R\_IS_{D_i}^{l_i}$; and
    \item Communications resources -  $R\_C_{D_i}^{l_i}$.
\end{itemize}

Users (clients) request infrastructure, service, and communication resources that are orchestrated by a network slicing software (NECOS, MATILDA, other) to create their sliced virtual network (SVN) as illustrated in Figure \ref{fig:SVNResources}.



The communication resources  $R\_C_{D_i}^{l_i}$ provide the interconnection of infrastructure and service resources $R\_IS_{D_i}^{l_i}$ for intradomain and inter-domain connections. As such, for the SVN modeling there are two distinct communication resources or communication slices (Figure \ref{fig:CommunicationSlices}):

\begin{itemize}
    \item Intradomain communication slices used between internal nodes of the domain:   $R\_C_{D_i[n_j, n_k]}^{l_i}$; and
    \item Interndomain communication slices used between domains:   $R\_C_{D_i,D_k}^{l_i}$
\end{itemize}

The communication slices are characterized by as set of parameters related to interdomain (Equation \ref{inter-domain}) and intradomain (Equation \ref{intra-domain}) communications:

\begin{eqnarray}\label{eq:PredictNextQ2} \label{inter-domain}
P\_RC_{D_i,D_j} = <B_{D_i,D_j}, L_{D_i,D_j}, Dl_{D_i,D_j}> 
\end{eqnarray}

\begin{eqnarray}\label{eq:PredictNextQ2} \label{intra-domain}
P\_RC_{n_i,n_j} = <B_{n_i,n_j}, L_{n_i,n_j}, Dl_{n_i,n_j}> 
\end{eqnarray}

Where:
\begin{itemize}
    \item $B_{D_i,D_j}$ is the available bandwidth between domains $D_i$ and $D_j$;
    \item $L_{D_i,D_j}$ is the packet loss between domains $D_i$ and $D_j$;
    \item $Dl_{D_i,D_j}$ is the delay between domains $D_i$ and $D_j$.
    \item $B_{n_i,n_j}$ is the available bandwidth between nodes $n_i$ and $n_j$ in a domain;
    \item $L_{n_i,n_j}$ is the packet loss between nodes $n_i$ and $n_j$ at a domain; and
    \item $Dl_{n_i,n_j}$ is the packet delay between nodes $n_i$ and $n_j$ at a domain.
\end{itemize}

Figures \ref{fig:SVNResources} and \ref{fig:CommunicationSlices} illustrate a generic view of the slicing process and related interdomain communications.  The network slicing infrastructure setup from the point of view of communication resources is as follows:

\begin{itemize}
    \item A set of domains ($D_i$);
    \item A single communication slice (configurable link or another communication resource) between domains; and
    \item A SDN switch (gateway) programmed to handling the interdomain packet routing among domains.
\end{itemize}

The interdomain slice communication parameters  $P\_RC_{D_k,D_j}^{l_i}$ are configured during the slicing commissioning phase, as proposed in the 3GPP network slicing reference architecture and model \cite{3gpp_3rd_2019}.


An SVN will require resources of distinct domains to be allocated end-to-end:

\begin{eqnarray}\label{eq:PredictNextQ2} \label{inter-domain}
SL_k^{D_i} = <R_i^{{D_i^{l_i}}}, R_j^{{D_i^{l_i}}},R_k^{D_y^{l_i}},...,R_z^{D_y^{l_i}}> 
\end{eqnarray}

\begin{figure}[htbp]
\begin{center}
\includegraphics[scale=0.5]{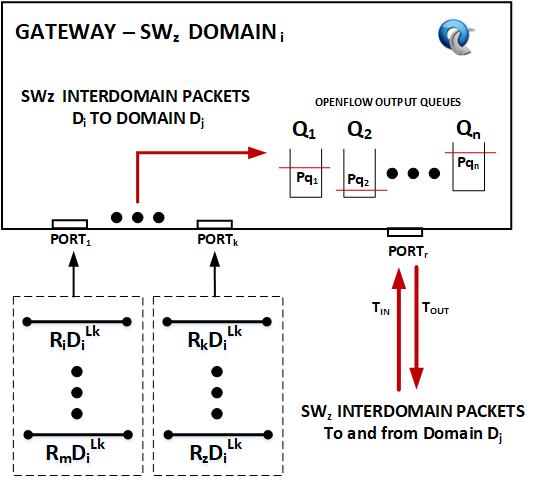}
\caption{Openflow Switch Handling Operation and Management Slicing Generated Packets}
\label{fig:SwitchSlicePackets}
\end{center}
\end{figure}

The communication slice modeling assumes that each domain contributes to a set of different resources that are located in various physical sites (domains).

\begin{figure*}[htbp]
\begin{center}
\includegraphics[scale=0.55]{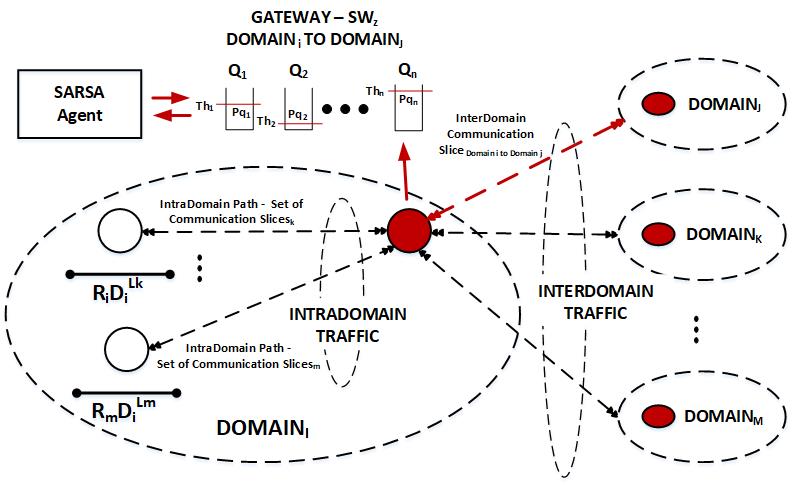}
\caption{Interdomain Communication Slice and Gateway at Domain i}
\label{fig:InterdomainCommunicationSlice}
\end{center}
\end{figure*}

The model is agnostic to the issue of traffic distinction between packets generated with the slices already instantiated (slice operation) and packets generated by the network slicing management software installed (orchestrator, resource marketplace, monitoring, others).

The slicing-related interdomain traffic between domains is handled by 
an SDN switch as illustrated in Figure \ref{fig:SwitchSlicePackets}.

In summary, the interdomain traffic at the gateway is composed of the packets generated (operation and management) by all resources belonging to the domain $D_i$ having as destination the domain $D_j$.

The slicing communication model assumes that domains have only one network connection together. In other words, the domains do not act as intermediate domains switching packets in the path to a destination domain.



For the interdomain packets at the gateway, the following definitions hold (Figure \ref{fig:SwitchSlicePackets}):

\begin{itemize}
    \item All packets belonging to a set of resources $R_i^{{D_i^{l_i}}}$ at domain $D_i$ with the same performance parameters constraint use a specific queue $Q_n$;
    \item $N$ switch queues handle the packet generated by the shareable resources at domain $D_i$;
    \item The switch queues have SDN resources control capabilities controlled by SDN Controllers  \cite{torres_sdnopenflow_2020} for resource control; 
    \item A priority is assigned to each output queue; and
    \item Each queue has a threshold level control parameter $P_{Q_n}$.
\end{itemize}

The priority and threshold level assigned to the queues are used to support for optimization (e.g. optimization controller as shown in the following section).


In summary, the model assumes that packets generated from any sliced resource with similar performance constraints are grouped in the same controlled queue in the gateway.




 


The following hypotheses are considered for the control of the intradomain packets and the gateway queues as highlighted in Figures \ref{fig:CommunicationSlices} and \ref{fig:InterdomainCommunicationSlice}):

\begin{itemize}
    \item Intradomain communications will be based on existing underlying communication technologies (MPLS LSPs connections, EON fiber slots, other);
    \item A gateway handles all the inbound and outbound interdomain traffics,
    \item In a domain, each node hosting sharing resources for the slicing creates a path to the gateway,
    \item Each path associated with a resource provided by a node is associated with a particular queue in the gateway.
\end{itemize}

The intradomain slice communication analytical model is not the focus of this paper, and these premises make clear its interrelation with the interdomain modeling and allows the independent modeling of it.

The optimization problem to solve here is the sharing of the communication resources between the different slices taking into account the QoS requirement of each slice. This means scheduling the packet originating from the different slices towards the different available queues in the gateway. This a complex engineering problem that is difficult to solve in an analytical way considering all the parameters that need to be taken into account. For that, we propose to investigate the use of a Reinforcement Learning SARSA agent which is explained in the following section.


\color{black}

\section{SARSA Agent to Optimize Resources Sharing} \label{sec:ProofOfConcept}

The interdomain communication slice model is now applied to the network slicing deployment setup illustrated in Figure \ref{fig:InterdomainCommunicationSlice} in which we have:

\begin{itemize}
    \item A multidomain slicing infrastructure with $n$ domains;
    \item A single communication slice between domains; and
    \item A SDN-capable switch (gateway) handling bidirectional interdomain packets between the domains.
\end{itemize}

In terms of the proof of concept, each interdomain communication slice has a reinforcement learning SARSA agent aiming to optimize the allocation of communication resources. The RL-SARSA agent acts during slice operation to dynamically keep performance parameters accordingly to management-defined objectives.





The interdomain slice communication parameters  ($P\_RC_{D_k,D_j}^{l_i}$) are configured during the slicing commissioning phase and are dynamically adjusted by the SARSA agent during the slice operation phase.

\subsection{SARSA Agent Model and Configuration} \label{sec:SARSAConfig}

The objective of the SARSA agent is to control the queue flushing transmission rates to preserve the performance parameters defined by the manager while sharing unused resources.

The slice communication queues ($Q_i$) are configured as follows:
\begin{itemize}
    \item Three queues corresponding to three performance parameters controlled by the agent;
    \item Each configured queue threshold ($Th_i$) corresponds to the performance parameter assigned to the queue and served to packets generated by sliced resources with this requirement, and
    \item  Each queue $Q_i$ has two states: below threshold (BT) and above threshold (AT).
\end{itemize}

The actions defined for the queues in the $AT$ state are to increase the transmission rate, reduce the transmission rate, and do nothing. Each executed state/action has a defined reward.

The SARSA agent and communication slice parameters and initial conditions for running are as follows:

\begin{itemize}
    \item Agent configuration parameters:
    \begin{itemize}
        \item Epsilon-greedy policy $\epsilon$ = 8$\%$;
        \item Learning rate $\alpha$ = 20$\%$; and
        \item Discount factor $\gamma$ = 80$\%$
    \end{itemize}
    \item Threshold limit (triggers agent action) = 50\%
    \item Agent actions: bandwidth increased or reduced by 10\%
    \item Maximum number of attempts = 500
    \item Queue priorities are: $p1$, $p2$ and $p3$ with $p1>p2>p3$.
\end{itemize}

SARSA Q-values are therefore updated based on the Equations \ref{eq:SARSA}:

\begin{eqnarray}\label{eq:SARSA}
\begin{split}
    Q(x_t,a_t) \leftarrow Q(x_t,a_t) + \alpha[r_{t+1} + \gamma Q(x_{t+1},a_{t+1}) \\ - Q(x_t, a_t)]
    \end{split}
\end{eqnarray}

\subsection{Implementation and tests} \label{sec:SARSAConfig}

The simulation environment was configured on a Linux (Ubuntu 22.04.1 LTS) Intel(R) Core(TM) i5-3470 CPU @ 3.20GHz desktop. The Visual Studio Code v.1.73.0 and Python v3.10.6 are used to execute the tests and the statistical analysis.

Each test run scenario has a minimum process cycle of $10^4$ packet production for each queue with a Poisson distribution.


The SARSA agent is called each time any queue reaches its configured threshold. The SARSA agent processes up to 500 episodes in search of a new configuration of the flushing bandwidth distribution among queues to keep buffer occupation in the configured threshold limit.

\subsection{The Slice Communication Evaluation Results} \label{sec:SliceCommunication}

A series of tests have been undertaken. It aims to overload the queues to evaluate the behavior of the agent. The three defined scenarios are the following:

\begin{itemize}
    \item Scenario 1 - One of the queues is overloaded;
    \item Scenario 2 - Two queues are overloaded; and
    \item Scenario 3 - All queues are overloaded.
\end{itemize}

The dynamics of the overloaded queues are configured as follows:
\begin{itemize}
    \item First set traffic 30\% above the queue defined limit for 10 minutes;
    \item Increase to 50\% above its defined limit for additional 10 minutes;
    \item Increase to 80\% above its defined limit for additional 10 minutes; and
    \item Increase to 100\% above its defined limit for additional 10 minutes.
\end{itemize}





\begin{figure*}[htbp]
\begin{center}
\includegraphics[scale=0.45]{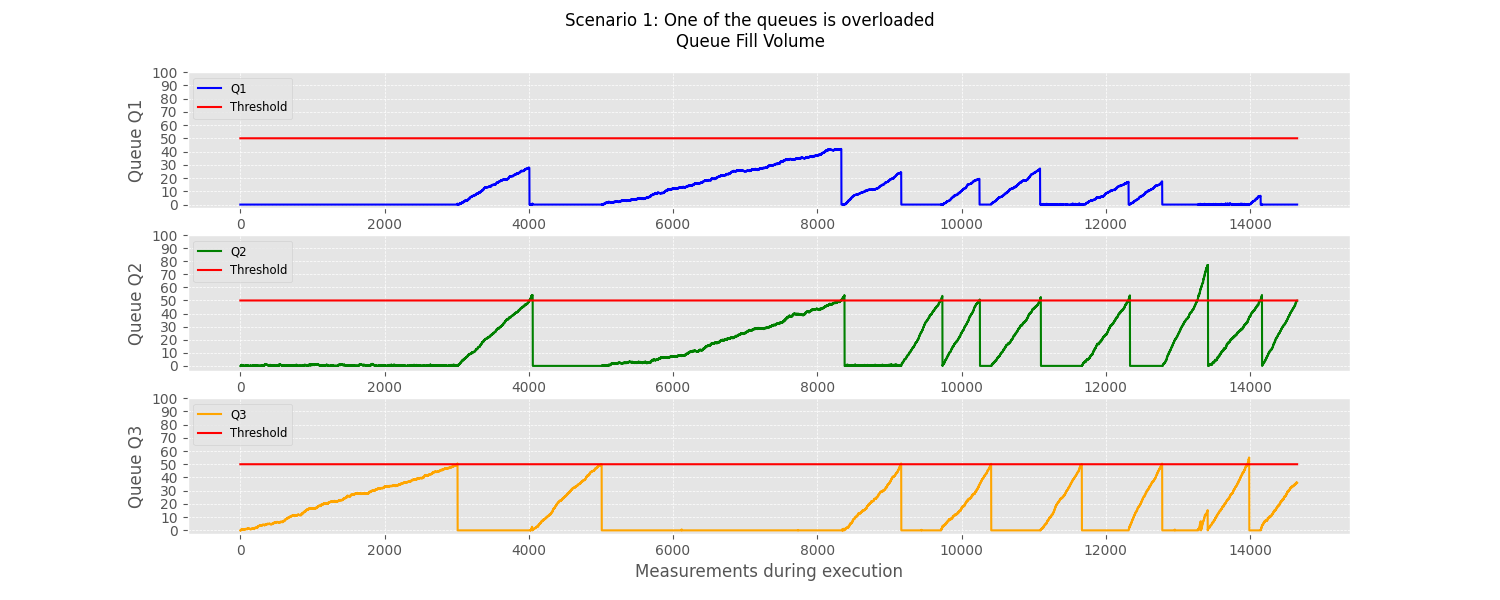}
\caption{Test Scenario 1 - One Gateway Queue Overloaded - Queue Size}
\label{fig:TestScenario1a}
\end{center}
\end{figure*}

\begin{figure*}[htbp]
\begin{center}
\includegraphics[scale=0.45]{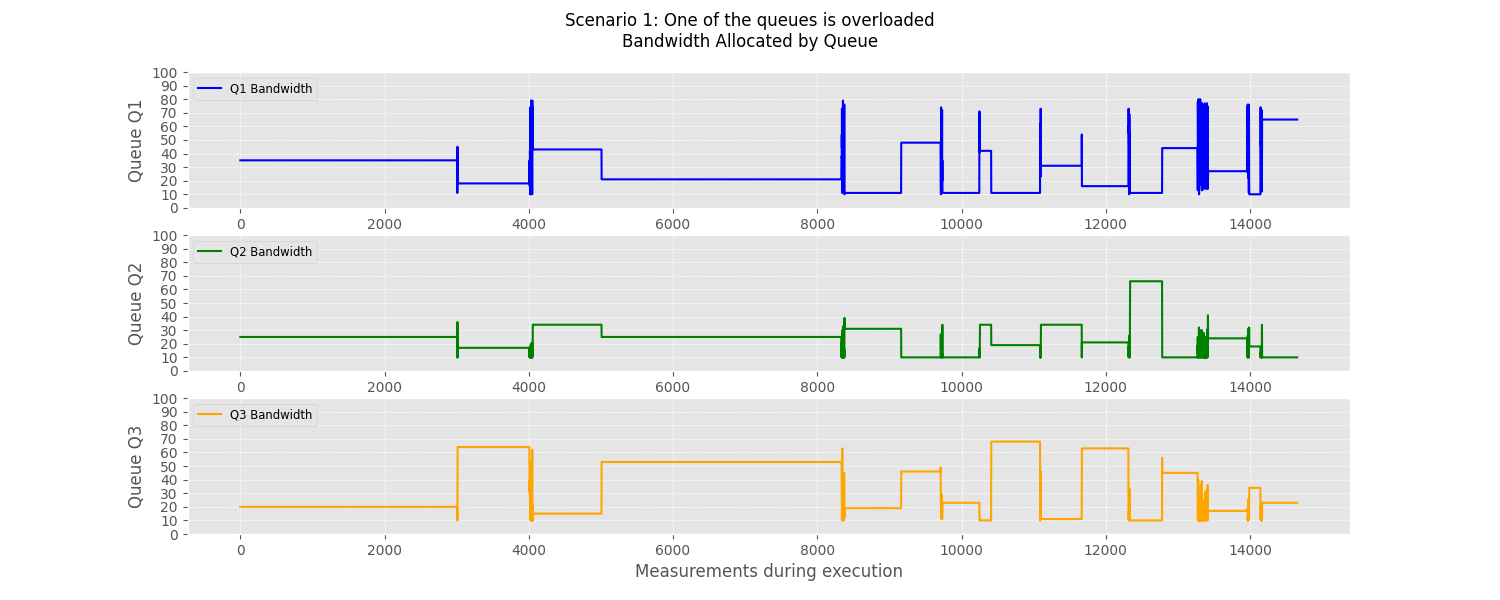}
\caption{Test Scenario 1 - One Gateway Queue Overloaded - Queue Flushing Rate}
\label{fig:TestScenario1b}
\end{center}
\end{figure*}

Figures \ref{fig:TestScenario1a} and \ref{fig:TestScenario1b} illustrate the SARSA agent's behavior for scenario one. Figure \ref{fig:TestScenario1a} plots the state of the queues while they are being saturated with overload traffic of packets. The queue transmission rate (flushing rate) configured by the SARSA agent is illustrated in Figure \ref{fig:TestScenario1b}. We observe that the total available bandwidth for the link is distributed and reconfigured among the queues according to the dynamic need to flush packets from a specific queue and keep queue occupation below the defined threshold.


\begin{figure*}[htbp]
\begin{center}
\includegraphics[scale=0.4]{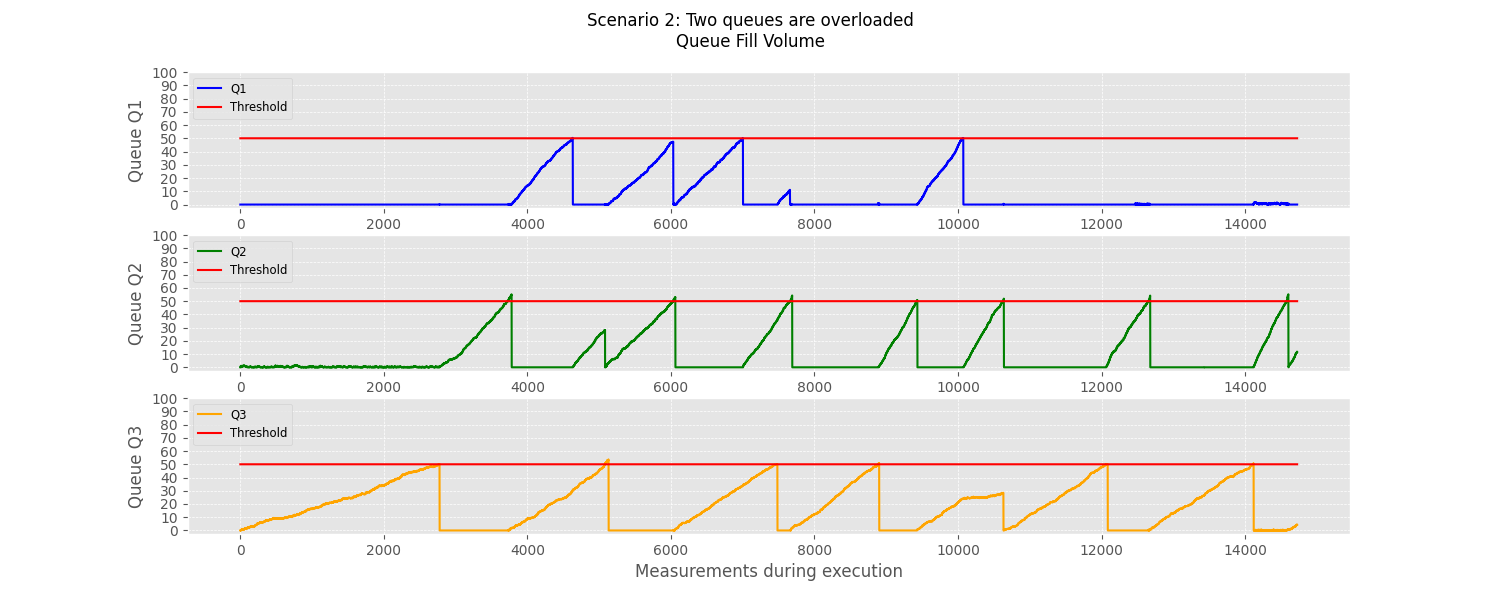}
\caption{Test Scenario 2 - Two Gateway Queues Overloaded - Queue Size}
\label{fig:TestScenario2}
\end{center}
\end{figure*}

For scenario two, the behavior of the SARSA agent is illustrated in Figure \ref{fig:TestScenario2}. In this scenario, two queues may overload, and, as observed in scenario one, the SARSA agent reconfigures the queue's transmission rate to keep buffer occupation below the defined threshold. The agent can deal with simultaneous overload for the simulation-defined parameters by keeping queue occupation as required.

\begin{figure*}[htbp]
\begin{center}
\includegraphics[scale=0.4]{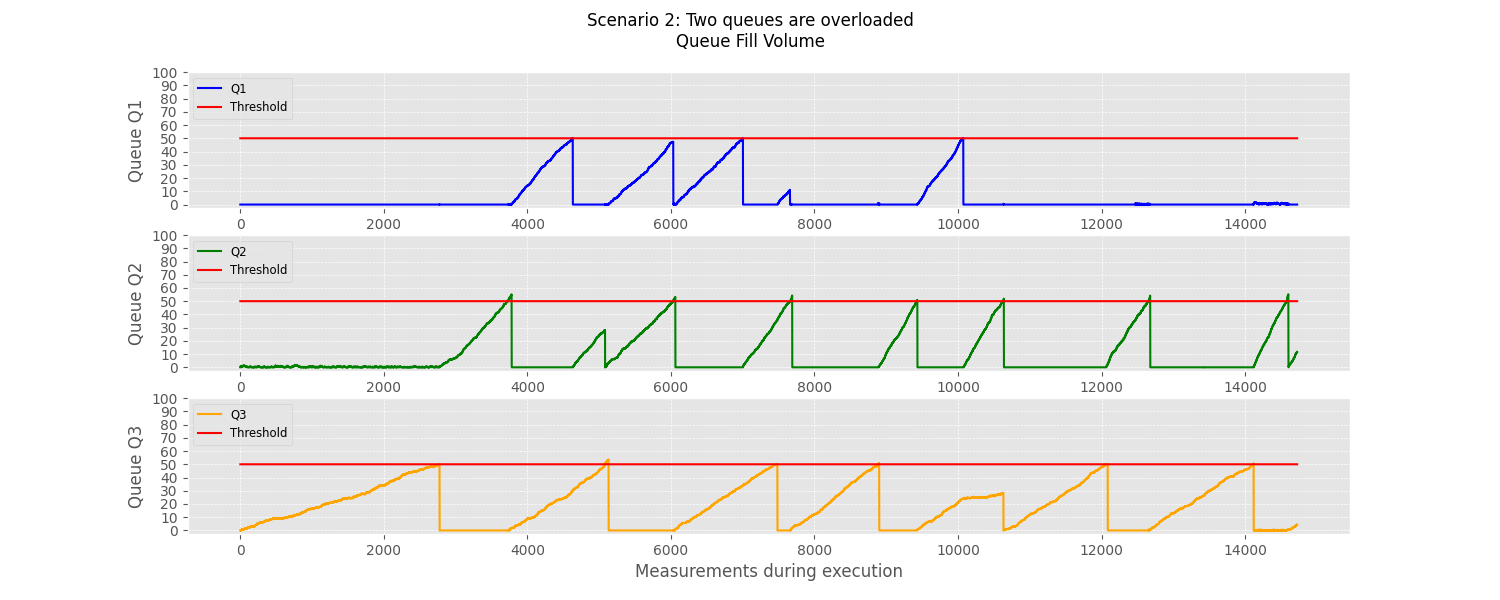}
\caption{Test Scenario 3 - Three Gateway Queues Overloaded - Queues Size}
\label{fig:TestScenario3}
\end{center}
\end{figure*}

Finally, the behavior of the SARSA agent for scenario 3 is illustrated in Figure \ref{fig:TestScenario3} and is equivalent to its behavior on scenario two.


\section{Final Considerations} \label{sec:Conclusion}



This paper presents a conceptual model of network slicing and present an analytical model to allocate communication resources between slide process. The conceptual model is along with a SARSA agent that optimize the allocation of communication resources among slices. The SARSA agent uses the conceptual model to formulate the required communication ressources of each slice. A proof of concept implementation of the SARSA agent aims to demonstrate that the SARSA agent contributes to dynamically adjusting and controlling the slice communication parameters between domains. The proposed conceptual model demonstrates the feasibility and ease of handling different types of communication resources for optimizing the communication slice. Future work includes the leverage of the conceptual model with the integration of intradomain and interdomain models and the new formulation of the distributed optimization problem to solve by a federation of SARSA agents.



\printbibliography

\end{document}